\def\cm#1{}

\documentstyle[pra,epsf,aps]{revtex}
\begin{document}
\title{{Addendum:  Strong-Coupling Behavior of $\phi^4$-Theories \\
and Critical Exponents [Phys. Rev. D {\bf 57}, 2264 (1998)]
}}
\author{Hagen Kleinert%
 \thanks{Email: kleinert@physik.fu-berlin.de ~~ URL:
http://www.physik.fu-berlin.de/\~{}kleinert ~~ Phone/Fax:
 0049 30 8383034 }}
\address{Institut f\"ur Theoretische Physik,\\
Freie Universit\"at Berlin, Arnimallee 14,
14195 Berlin, Germany}
\maketitle
\begin{abstract}
The graphical extrapolation procedure to infinite order
of variational perturbation theory in a recent calculation of critical exponents
of three-dimensional $\phi^4$-theories
at infinite couplings is improved
by another way of plotting the results.
\end{abstract}

%
\noindent
In a recent calculation of critical exponents
of three-dimensional
$\phi^4$-theories
which was based on the strong bare-coupling limit
of variational perturbation theory \cite{kl},
the final results were obtained by extrapolating the
approximate exponents
of order 2,3,4,5, and 6 to order $N\rightarrow \infty$
using the theoretically
calculated large-$N$ behavior const$+e^{-cN^{1- \omega}}$,
where $ \omega$ is the critical exponent of the approach to scaling.
The plots showed how the  exponents approach
their limiting values as functions of $N$.
The purpose of this note is
to point out
that
the extrapolation
can be improved
by plotting the same
exponents
against the variable $x=e^{-cN^{1- \omega}}$
rather than $N$, and varying the parameter $c$ until
the points merge approximately into a straight line.
Its intercept with the vertical axis yields the
desired extrapolated critical exponent.

The paper \cite{kl}
also contains misprints in
some figure labels
which are corrected in the corresponding present
figure headings. They are also corrected in the paper
on the APS eprint server (also available from
http://www.physik.fu-berlin.de/\~{}kleinert\/klein\_re3\#257).

A clarifying remark is in order concerning the
extremizing of Eq.~(8). The
parameter $c_N$
was expanded into a power series in $g_0^{-q/2}$,
after which the energy was extremized in the coefficients.
The procedure
requires a reexpansion
of the coefficients $b_i(c_N)$
as descibed in the earlier paper
\cite{int} [see Eq.~(11) and the subsequent paragraph
in that paper],
a fact which was not
clearly stated in \cite{kl}.

Let us also point out that
the theory can be applied
to the perturbation expansions
of the renormalization constants
in $4- \epsilon$ dimensions
\cite{klep},
where resummed $ \epsilon$-expansions are obtained
without the
traditional use of renormalization group
and Pade\'e-Borel methods.

Finally, the first reference in
\cite{kl} was incomplete
and should read as
shown in
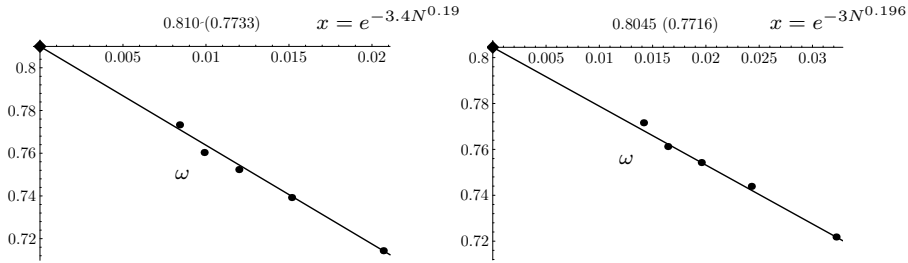
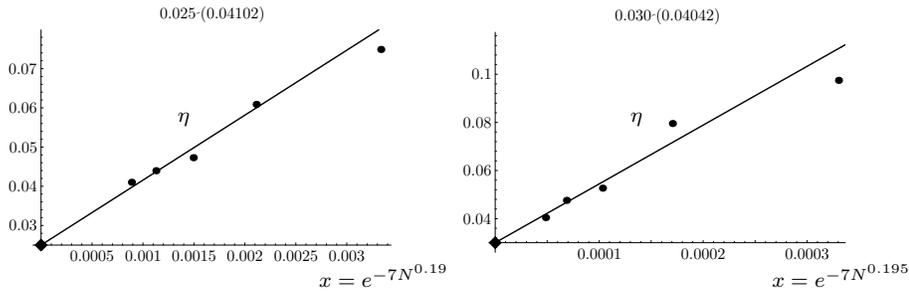
\cite{kl2}.

~\\
\begin{figure}[tbhp]
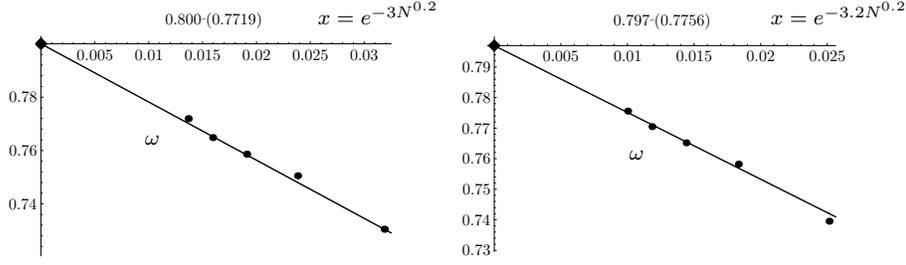

~~~~~~\input cromnc0f.tps~~~
\input cromnc1f.tps~~~\\~~\\ ~~\\  ~~\\~~\\~~\\
\phantom{xxxxx}\,\,\input cromnc2f.tps   ~~~
\input cromnc3f.tps   ~~~
~\\~\\~\\      ~\\
\caption[]{Behavior
of the strong-coupling
values of the critical exponent of approach to scaling
$ \omega$
with increasing orders $N=2,3,4,5,6$
in variational perturbation theory
for
O($n$)-symmetric theories with $n=0,~1,~2,~3,~\dots~$.
The plot is against  $x=e^{-cN^{1- \omega}}$
with $c$ chosen such that the points merge into a straight line.
Its
intercept with the vertical axis
yields the extrapolated result of infinite order.
The numbers $n$ correspond to different universality
classes ($n=0,1,2,3$ fo percolating systems,
Ising magnets, superfluid Helium, and the classical Heisenberg model).
The numbers on top display the limiting value at the intercept, as well as
 the last
calculated approximation $ \omega_6$ in parentheses.
This number is found also in Table 1 of Ref.~\cite{kl}.
}
\label{crom1}\end{figure}
\begin{figure}[tbhp]
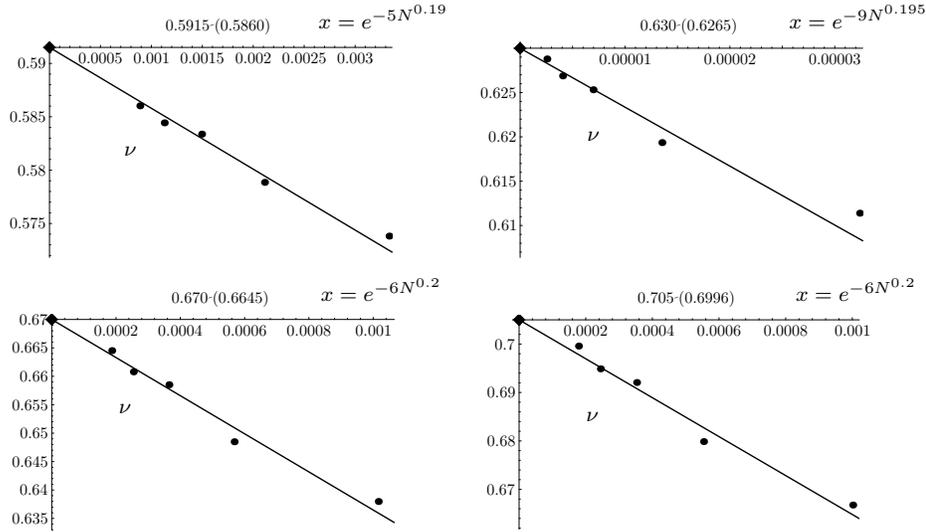

~~~~~~\input crnunc0f.tps~~~
~~\input crnunc1f.tps~~~\\~~\\ ~~\\  ~~\\~~\\
\phantom{xxxxx.}\,\input crnunc2f.tps   ~\,~~~
~\input crnunc3f.tps   ~~~
~\\~\\~\\      ~\\
\caption[]{Plot analogous
to Fig.~\ref{crom1} of the
critical exponents
$ \nu$
for
increasing orders $N=2,3,4,5,6$ in variational perturbation theory,
illustrating the extrapolation procedure
to $N\rightarrow \infty $ at the intercept with the vertical axis, for $n=0,~1,~2,~3$.
The numbers on top display the limiting value at the intercept, as well as
 the last
calculated approximation $ \nu_6$ in parentheses.
The latter were misprinted in \cite{kl}.
This number is found also in Table 1 of Ref.~\cite{kl}.
}
\label{nu1}\end{figure}
\begin{figure}[tbhp]
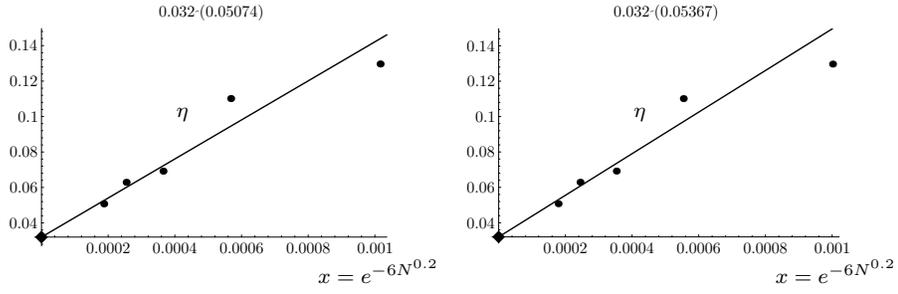

~~~~~~~~~\input cretnc0f.tps~~~
\input cretnc1f.tps~~~\\~~\\ ~~\\  ~~\\~~\\
\phantom{xx,.xxxx}\,\input cretnc2f.tps   ~~~
\,\input cretnc3f.tps   ~~~
~\\~\\~\\      ~\\
\caption[]{
Plot analogous
to Figs.~\ref{crom1} and \ref{nu1} of the
critical exponents
$ \eta=\eta(\infty)$
for
increasing orders $N=2,3,4,5,6$ in variational perturbation theory,
illustrating the extrapolation procedure
to $N\rightarrow \infty $ at the intercept with the vertical axis, for $n=0,~1,~2,~3$.
The approximations $ \eta_N$ are obtained from
 $ \gamma_N$ via the scaling relation $ \eta_N=2- \gamma_N/ \gamma$.
The numbers on top display the limiting value at the intercept, as well as
 the last
calculated approximation $ \eta_6$ in parentheses.
The latter were misprinted in \cite{kl}.
}
\label{et1}\end{figure}
\pagebreak
~\\

\begin{figure}[tbhp]
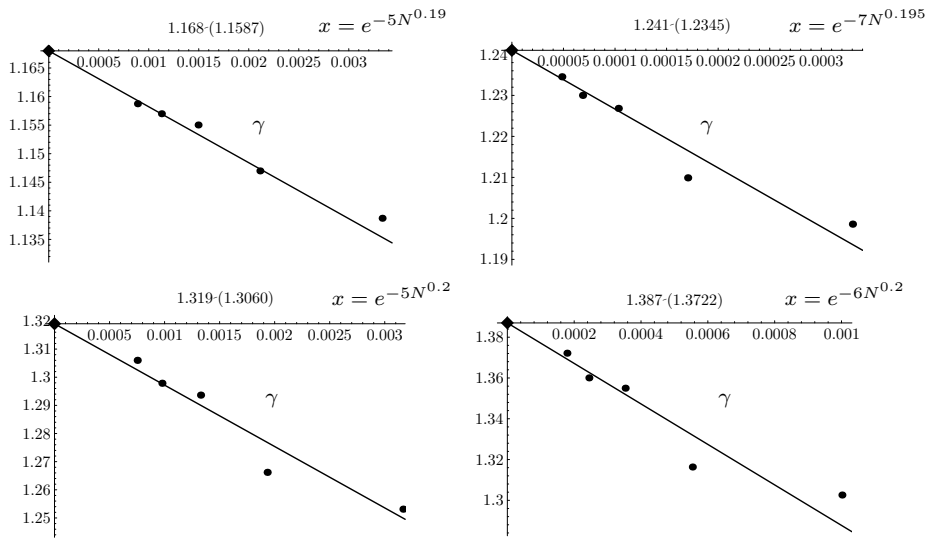

~~~~~~\input crganc0f.tps~~~
~~\input crganc1f.tps~~~\\~~\\ ~~\\  ~~\\~~\\
\phantom{xxxxxx}\,\,\input crganc2f.tps   ~~~
\input crganc3f.tps   ~~~
~\\~\\~\\      ~\\
\caption[]{
Plot analogous
to Figs.~\ref{crom1}, \ref{nu1}, and \ref{et1}
of the
critical exponents
$ \gamma=\gamma(\infty)$
for
increasing $N=2,3,4,5,6$ in variational perturbation theory,
illustrating the extrapolation procedure
to $N\rightarrow \infty $ at the intercept with the vertical axis, for $n=0,~1,~2,~3$.
This number is found also in Table 1 of Ref.~\cite{kl}.
}
\label{ga1}\end{figure}

~\\~\\Acknowledgement\\
The author is grateful to Dr. V. Schulte-Frohlinde
for many very useful discussions.

\end{document}